\documentclass[notoc]{JHEP3} 

\usepackage{amsmath}
\usepackage{fleqn} 
\usepackage{indentfirst} 
\usepackage{mathrsfs}
\usepackage{accents}
\usepackage{HEP}

\newcommand\fverb{\setbox\pippobox=\hbox\bgroup\verb}
\newcommand\fverbdo{\egroup\medskip\noindent%
			\fbox{\unhbox\pippobox}\ }
\newcommand\fverbit{\egroup\item[\fbox{\unhbox\pippobox}]}
\newbox\pippobox
\def\D{\mathrm{d}} 
\def\DS{\mathrm{D}}
\def\Int{\int\limits}
\def\a{{\alpha}}

\def\b{{\beta}}

\def\t{{\theta}}

\def\<{\left\langle}
\def\>{\right\rangle}
\title{Coupling the Superstring to a D-Brane Ramond-Ramond Background}

\author{Stefan Antusch\\
Physik-Department T30, 
Technische Universit\"{a}t M\"{u}nchen\\ 
James-Franck-Stra{\ss}e,
85748 Garching, Germany\\
E-mail: \email{santusch@ph.tum.de} 
}

\preprint{{\small TUM-HEP-428/01}}

\abstract{We propose a new approach for coupling the 
type II superstring to the Ramond-Ramond 
background of D-branes in the RNS formalism,
alternative to introducing RR vertex operators. 
It is based on the mixing 
between Ramond-Ramond $p$-form 
excitations in the closed string spectrum and
transversally polarized
excitations in the open string spectrum.
}


\begin{document} 

\section{Introduction}
D-branes \cite{Polchinski:1996na} have played a crucial role in the recent 
development of string theory. 
For example, they have been important for realizing the duality relations 
between the five consistent string theories and 
have given rise to phenomenologically interesting models. 
For some of them and for the AdS/CFT correspondence 
\cite{Maldacena:1998re,Gubser:1998bc,Witten:1998qj}, 
it is desirable to improve the understanding of 
string theory in the presence of the background fields of D$p$-branes. 
The major difficulty is the handling of the Ramond-Ramond (RR) background. 
In the Ramond-Neveu-Schwarz (RNS) 
formalism, the usual method for coupling the string to a background field 
fails as 
superconformal invariance is spoiled if the RR vertex operators are used to
deform the action. 

There have been many approaches to this problem in recent years, for example 
\cite{Berenstein:2001ip,Metsaev:1998it,Berkovits:1999im,Polyakov:1999pm,Naik:2001pr,Berkovits:2001wm},
using different formalisms.
In the RNS formalism, the interaction Lagrangian with RR vertex operators  
has been investigated in \cite{Berenstein:2001ip}. 
The approaches 
\cite{Metsaev:1998it,Berkovits:1999im,Polyakov:1999pm,Naik:2001pr,Berkovits:2001wm} 
are either based on the Green-Schwarz (GS) formalism or on a
mixture between the RNS and the GS formalism, such as the hybrid formalism.
In \cite{Metsaev:1998it}, the coset supermanifold
$\mbox{SU}'(2,2|4)/(\mbox{SO(4,1)}\otimes \mbox{SO(5)})$ 
is taken as target space for
the  GS superstring in the $AdS_5 \otimes S^5$ RR background. 
The coupling to the RR
background found this way agrees to some extent  
with the one constructed by performing
double dimensional reduction and a T-duality transformation of the 
11-dimensional supermembrane action coupled to the
supergravity background \cite{Cvetic:2000zs}.
The approach \cite{Berkovits:1999im} in the hybrid formalism  
is based on the topological string theory in
RNS variables and partly introduces GS variables. The RR
vertex operators can then be used to deform the action of the 
free superstring to
achieve the coupling to the RR background.

In this letter we propose a new way for coupling the superstring 
to the RR background
of a flat D-brane in the RNS formalism, 
alternative to introducing RR vertex operators.

\section{Coupling of 1-Forms to the RNS Superstring} \label{sec:1forms}
In the bosonic string theory, the coupling of a 1-form background field is
given by integrating the pullback of the 1-form $A^{(1)}$ 
over the string boundary $\partial \Sigma$. With the embedding $X$ 
and the pullback $X^*$, 
it can be written as
\begin{eqnarray}\label{eq:bos1form}
  \Int_{\partial \Sigma}\, X^* \,A^{(1)} &=&   \Int_{\partial
  \Sigma}\,A^{(1)}(X(z,\overline z\,)) = \Int_{\partial
  \Sigma}\, A^{(1)}_\mu (X(z,\overline z\,)) \,\D X^\mu (z,\overline z\,) 
  \nonumber\\
  &=&  \Int_{\partial
  \Sigma}\, A^{(1)}_\mu (X(z,\overline z\,))\,\{ \partial X^\mu \D z 
  + \overline \partial  X^\mu
  \D \overline z\,\}\; .
\end{eqnarray}
This has to be modified if D$p$-branes are present. 
In Cartesian coordinates and
with $i \!\in \!\{ 0, \dots, p\}$ denoting indices parallel 
and $t \!\in\!\{ p\!+\!1, \dots, 9\}$ indices transverse to the $p$-brane
we obtain the proper
coupling by T-dualizing equation (\ref{eq:bos1form}), 
\begin{eqnarray}\label{eq:bos1formTd}
\Int_{\partial\Sigma}  \,A^{(1)}_{i}( X) \,\{ \partial { X}^i
\D z +  \overline \partial { X}^i  \D \overline z \}  + 
\Int_{\partial\Sigma}  \,\Phi_{t}( X) \,\{ \partial { X}^t
\D z -  \overline \partial { X}^t  \D \overline z \} \; ,
\end{eqnarray}
where we have defined $\Phi_{t}:=A^{(1)}_{t}$.
Supersymmetrizing equation (\ref{eq:bos1formTd}), i.e. substituting the 
components $X^\mu$ of the embedding by the superfields 
${\boldsymbol X}^\mu = X^\mu +  \t \psi^\mu+  \overline \t 
\,\widetilde \psi^\mu +  \overline \t \t F^\mu$, the Dolbeaut operators 
$\partial$ and $\overline \partial\,$ 
by $\DS =\partial_\t + \t \partial$ and 
$\overline \DS \,=\partial_{\overline \t}+
\overline \t \,\overline \partial\, $, and
adding the integration over the complex fermionic coordinates 
$\overline \t $ and $ \t$ yields the
coupling
\begin{eqnarray}\label{eq:bos1formTdRNS}
\lefteqn{\Int_{\partial\Sigma} A^{(1)}_{i}(\boldsymbol X)
 \,\{ \DS  {\boldsymbol X} ^i \,\D z \,\D \theta  +
\overline \DS  {\boldsymbol X}^i \,\D \overline z\, \D \overline \theta\,\} }\nonumber \\
&+& \Int_{\partial\Sigma}\Phi_t(\boldsymbol X)
 \,\{ \DS  {\boldsymbol X} ^t \,\D z \,\D \theta  -
\overline \DS  {\boldsymbol X}^t \,\D \overline z\, \D \overline \theta\,\} 
\end{eqnarray}   
of a 1-form background field to the superstring in the RNS formalism.

\section{Construction of the Interaction Lagrangian}\label{sec:CouplinTerm}
The RR background of a D$p$-brane is given by a $p\!+\!1$-form potential
$C^{(p+1)}$. 
The translation symmetries which are unbroken by the flat D$p$-brane
are generated by the
$p\!+\!1$ vector fields  
$V_0 := \frac{\partial}{\partial X^0}, 
\dots, V_p :=\frac{\partial}{\partial X^p}$.  
With the interior products
$i_{V_0}, \dots, i_{V_p}$ and the exterior derivative $\D $ as tools, 
from $C^{(p+1)}$ an interesting 1-form,
\begin{eqnarray}
i_{V_0} \cdots i_{V_p} \,\D  C^{(p+1)} = i_{V_0} \cdots i_{V_p} \,F^{(p+2)} \; ,
\end{eqnarray}  
can be constructed. $F^{(p+2)}$ is the $p\!+\!2$-form field strength. 
The pullback of this 1-form can
now be geometrically coupled to the string boundary as in 
section \ref{sec:1forms} or equivalently the closed
2-form $\D  \,i_{V_0} \cdots i_{V_p} \,\D  C^{(p+1)}$ can be 
coupled to the string itself. Supersymmetrizing and T-dualizing,
 as in section \ref{sec:1forms}, yields
\begin{eqnarray}\label{eq:LagrIntRR}
\mathscr{L}_\mathrm{int} = \Int_{\partial\Sigma} \,   F _{0\dots p t} (\boldsymbol X) 
\, \{\DS  {\boldsymbol X}^t \D z\,\D \t - \overline \DS  {\boldsymbol X}^t
 \D \overline z\,\D \overline \t     \}         \: .  
\end{eqnarray} 
We propose that this is a viable interaction Lagrangian for the superstring in
the RR background of a flat D$p$-brane. 

In general,
interaction Lagrangians can be constructed using the
vertex operators corresponding to the desired background field. 
In the RNS formalism, RR vertex operators contain spin fields and this method
leads to the loss of superconformal invariance. 
On the other hand, the 
interaction Lagrangian proposed 
in equation (\ref{eq:LagrIntRR}) can be viewed as being constructed
from a vertex operator 
for an open
string state polarized transversally to the D$p$-brane. 
This can be seen more explicitly
by writing (\ref{eq:LagrIntRR}) in component fields,
\begin{eqnarray} \label{eq:LagrIntRRcomp}
\mathscr{L}_\mathrm{int} = \Int_{\partial\Sigma} \D z\:  F_{0\dots p t} (X(z,\overline z\,)) \, 
 [\partial X^t (z) +
 (2\accentset{\leftarrow}{\partial_i}\psi^i (z)) \psi^t (z)] \; ,
\end{eqnarray}
using the notation $F _{0\dots p t} (X)
\accentset{\leftarrow}{\partial_i} \,:=\, \partial_i F _{0\dots p t} (X)$. 
Thus, for the proposed interaction Lagrangian of equation (\ref{eq:LagrIntRR}) 
to be appropriate, 
the vertex operator for the 
transversally polarized open string should be 
a proper substitute for
the RR vertex operator.

\section{Mixing between States of the Open and the Closed String}
\label{sec:RRvsOpen}
It is known that a mixing between excitations in the open and 
closed string spectrum can occur. 
In order to investigate the relation between closed strings in RR states 
and open strings polarized transversally to a D-brane, we therefore 
consider the disk amplitude\footnote{A recent detailed calculation 
of more general disk 
amplitudes with RR vertex operators in the 
$(-\frac{1}{2},-\frac{3}{2})$
ghost picture can be found in \cite{Liu:2001qa}.} with the vertex operators
\begin{eqnarray}
V^{\mathrm{RR}} {(q)}&=& e^{-\frac{1}{2}\phi}S^{\alpha}(z)\, 
 f_{ \mu_1 \dots \mu_{p+1}} \Gamma_{\a \b}^{\mu_1 \dots \mu_{p+1}} \, 
  e^{-\frac{1}{2}\overline \phi\,}\overline S\,^{\beta}(\overline z\,
  )\,e^{i q X(z,\overline z\,) }\; , \\
V^{\mathrm{open}} {(k)}&=& \xi_t  \,e^{-\phi} \psi^t\,  e^{i2k_\parallel X} (w)\; ,   
\end{eqnarray}
for the RR state in the $(-\frac{1}{2},-\frac{1}{2})$ ghost 
picture
($S^{\alpha}$ are the spin fields) and 
the transversally polarized open string state 
in the $(-1)$ ghost picture, respectively. 
Using the doubling trick, this yields
\begin{eqnarray}\label{eq:MixingRR-Open}
\mathscr{A} (V^{\mathrm{RR}}(q) V^{\mathrm{open}}(k))_\mathrm{D}  &=&\<  c(z)c(\overline z\,)
V^{\mathrm{RR}}_{(q)} (z,\overline z\,) \,
c(w)V^{\mathrm{open}}_{(k)}(w)\>_\mathrm{D} \nonumber \\
&\sim& f_{0\dots p t} \, \xi^{t}  \,
\delta (2 q_\parallel + 2 k_\parallel ) \; .
\end{eqnarray}
Thus, open strings with a transverse polarization
mix with closed strings in RR states with the  
polarization $f_{0\dots p t}\not=0$. 
Therefore, the open string vertex operators can indeed act as a substitute 
for the RR vertex operators, 
which is an argument in favor of  
the validity of the interaction
Lagrangian (\ref{eq:LagrIntRR}).

\section{Conclusions}
We have proposed an interaction 
Lagrangian  for the
superstring in the RR background of a flat D$p$-brane, 
alternative to introducing the RR vertex operator.
It has been
constructed using the translation symmetries which are unbroken by the
D$p$-brane and can be 
viewed as effecting the coupling to  
the RR background by means of transversally polarized
open strings. 
These states of the
open string mix with the RR states of the closed string
corresponding to the background of the D$p$-brane, as shown in equation
(\ref{eq:MixingRR-Open}),  
providing the main argument for the proposed interaction Lagrangian. 

The starting point for further investigations, the proposed complete action for the type II superstring 
in the RR background of a
D-brane including the usual gravitational background $G_{\mu\nu}$ and the
dilaton $\phi$, is thus given by
\begin{eqnarray}\label{eq:ComplRRinRNS}
S &=& 
\frac{1}{2}\,\Int_{\Sigma} \D ^2z \,\D ^2\theta \:  G_{\mu\nu}(\boldsymbol X) \,
 \overline \DS {\boldsymbol X}^\mu 
 \DS {\boldsymbol X}^\nu  + \Int_{\Sigma} 
 \D ^2z \, \phi (X) R^{(2)}  \nonumber \\
&+&
\Int_{\partial\Sigma} \,   F _{0\dots p t} (\boldsymbol X) 
\, \{\DS {\boldsymbol X}^t \,\D z\,\D \t - 
\overline \DS {\boldsymbol X}^t \,\D \overline z\,\D \overline \t \} \; . 
\end{eqnarray} 

\section*{Acknowledgements}
The author would like to thank I.~Sachs, C.~Scrucca, E.~Scheidegger 
and  S.~Theisen
for useful discussions.
This work was supported by the 
``Sonderforschungsbereich~375 f\"ur Astro-Teilchenphysik der 
Deutschen Forschungsgemeinschaft''.

\end{document}